\documentclass{PoS}
\usepackage[authoryear]{natbib}
\bibpunct{(}{)}{;}{a}{}{,}

\usepackage{comment}
\usepackage{amsmath,amssymb}

\title{Measuring magnetism in the Milky Way with the Square Kilometre Array}

\ShortTitle{Magnetism in the Milky Way with the SKA}

\author{Marijke Haverkorn$^{1,2}$\footnote{On behalf of the SKA
    Cosmic Magnetism Working Group}, Takuya Akahori$^3$, Ettore
  Carretti$^4$, Katia Ferri\`ere$^5$, Peter Frick$^6$, Bryan Gaensler$^{3,7}$, George Heald$^8$, Melanie
  Johnston-Hollitt$^{9}$, David Jones$^1$, Tom Landecker$^{10}$,
  \speaker{Sui Ann Mao}$^{11}$, Aris
  Noutsos$^{11}$, Niels Oppermann$^{12}$, Wolfgang Reich$^{11}$, Timothy
  Robishaw$^{10}$, Anna Scaife$^{13}$, Dominic Schnitzeler$^{11}$, Rodion
  Stepanov$^{6,14}$, Xiaohui Sun$^3$, Russ Taylor$^{15,16}$
\\
$^1$Radboud University; $^2$Leiden University; $^3$The University of
Sydney; $^4$CSIRO Astronomy and Space Science; $^5$IRAP, Universit\'e de Toulouse, CNRS; $^6$Institute of Continuous Media Mechanics ;
$^7$ARC Centre of Excellence for All-Sky Astrophysics (CAASTRO);
$^8$Netherlands Institute for Radio Astronomy (ASTRON); $^{9}$Victoria
University of Wellington; $^{10}$Dominion Radio Astrophysical
Observatory; $^{11}$Max-Planck-Institut f\"ur Radioastronomie;

$^{12}$CITA; $^{13}$University of Southampton; $^{14}$Perm National
Research Polytechnic University;$^{15}$University of Cape
Town;$^{16}$University of the Western Cape\\
E-mail:\email{m.haverkorn@astro.ru.nl}
}
\abstract{

  Magnetic fields in the Milky Way are present on a wide variety of
  sizes and strengths, influencing many processes in the Galactic
  ecosystem such as star formation, gas dynamics, jets, and evolution
  of supernova remnants or pulsar wind nebulae. Observation methods
  are complex and indirect; the most used of these are a grid of
  rotation measures of unresolved polarized extragalactic sources, and
  broadband polarimetry of diffuse emission. Current studies of
  magnetic fields in the Milky Way reveal a global spiral magnetic
  field with a significant turbulent component; the limited sample of
  magnetic field measurements in discrete objects such as supernova
  remnants and HII regions shows a wide variety in field
  configurations; a few detections of magnetic fields in Young Stellar
  Object jets have been published; and the magnetic field structure in
  the Galactic Center is still under debate.

  The SKA will unravel the 3D structure and configurations of magnetic
  fields in the Milky Way on sub-parsec to galaxy scales,
    including field structure in the Galactic Center. The global
  configuration of the Milky Way disk magnetic field, probed through
  pulsar RMs, will resolve controversy about reversals in the Galactic
  plane. Characteristics of interstellar turbulence can be determined
  from the grid of background RMs. We expect to learn to understand
  magnetic field structures in protostellar jets, supernova remnants,
  and other discrete sources, due to the vast increase in sample sizes
  possible with the SKA. This knowledge of magnetic fields in the
  Milky Way will not only be crucial in understanding of the evolution
  and interaction of Galactic structures, but will also help to define
  and remove Galactic foregrounds for a multitude of extragalactic and
  cosmological studies.  }

\FullConference{
Advancing Astrophysics with the Square Kilometre Array\\
June 8-13, 2014\\
Giardini Naxos, Italy}

\begin{document}

\section{Introduction}

In the past years, cosmic magnetism science has transformed from a
niche to a fast growing and exciting field of research. This is largely due
to improvements in computational power, and enhanced observational
capabilities. The latter led to the presence of extensive new data
sets like the Canadian \citep{Landeckeretal10} and Southern
\citep{Haverkornetal06a} Galactic Plane Surveys, the Effelsberg 1.4~GHz Medium
Latitude Survey \citep[EMLS,][]{Reichetal04}, the Urumqi 6~cm survey
\citep{Sunetal07}, and the S-band Polarization All-Sky Survey (S-PASS)
\citep{Carrettietal14}; and ongoing surveys like GALFACTS
\citep{TaylorSalter10} or GMIMS \citep{Wollebenetal09}. Unraveling the
structure and strength of magnetic fields in the Universe is such a
complex and involved system, however, that much is still under debate
or plainly unknown.

Studies of magnetic fields in galaxies are at the center of many open
questions regarding cosmic magnetism.  Current galactic magnetic
fields may originate in the primordial universe, possibly from
magnetizing topological defects, from early stellar magnetic fields
\citep{Widrow02} or from magnetic fields in the cosmic web
\citep{Sofueetal10}. Seed fields may also be generated in protogalaxies
by e.g.\ the Biermann mechanism in the first supernova remnants
\citep{Hanayamaetal05}, the Weibel instability \citep{Lazaretal09}, or
from plasma fluctuations \citep{SchlickeiserFelten13}.
Magnetic fields in the intracluster and intergalactic medium may have
originated from outflows of galaxies \citep{Donnertetal09}.  On smaller
scales, galactic magnetic fields interact with supernova remnants, and
help shape fields in star formation regions \citep{Fishetal03,
  Greenetal12}.

The Milky Way is an ideal testbed for many galactic magnetic field
studies due to its proximity. We have the most detailed observations
and understanding of the multi-phase interstellar environment in the
Milky Way to help interpret magnetic field studies. In particular, for
magnetic fields in discrete objects and on smaller scales
  ($\lesssim 1$~pc), the Milky Way is the best location where these
  can be studied in detail (although parsec-scales become
    accessible in the Magellanic Clouds as well, see
    \citet{becketal15}). The global configuration of the magnetic
field and characterization of magnetic fields in the gaseous halo is
difficult to determine from our location within the Milky Way, and is
more easily studied in other (face-on) spiral galaxies. However, the
Milky Way provides the opportunity of studying large-scale reversals
in this global configuration of the field in unique detail.

In addition, the Milky Way is a significant foreground for a wide
variety of extragalactic studies. The magnetic fields in the Milky Way
disk and halo provide a characteristic radio-polarimetric pattern,
which needs to be understood to enable study of e.g.\ the Epoch of
Reionization, inflation theory through the Cosmic Microwave Background
B-mode polarization, or magnetization of the cosmic web. Also,
knowledge of the structure of the Galactic magnetic field is
imperative to trace back Ultra-High Energy Cosmic Rays to their
sources, and to understand the multi-phase, multi-component
interstellar medium (ISM).

In the Square Kilometre Array (SKA) Science Case written in 2004,
\cite{BeckGaensler04} described SKA observations with magnetic fields
in the Milky Way and nearby galaxies, and stated the importance of an
{\em all-sky Rotation Measure (RM) Grid} of compact extragalactic radio
sources. In the updated SKA Science Case for magnetic fields in the
Milky Way presented in this chapter, we re-emphasize the importance of
the RM Grid, but in addition explain the necessity of broadband
polarimetry of diffuse synchrotron emission, enabling Faraday
Tomography through RM synthesis \citep{BrentjensDeBruyn05}.

Section~\ref{s:MWBfields} details the current status of research in
magnetic fields in the Milky Way, from the global configuration and
turbulent fields, to fields in discrete objects and in the Galactic
Center. In Section~\ref{s:ska}, we describe the Galactic magnetism
studies that will become possible with the
SKA. Subsection~\ref{s:methods} details the two methods used, viz.\ the
RM Grid and Faraday Tomography, whereas in the next three subsections
we present the possibilities with SKA in phase 1 Early Science, SKA in
phase 1 and SKA in phase 2. 
.

\section{Magnetic Fields in the Milky Way}
\label{s:MWBfields}

Galactic magnetic fields are amplified and maintained by magnetic
dynamo action, through conversion of energy from e.g.\ differential
rotation, turbulence and/or cosmic rays into magnetic energy. Regular
fields may be amplified by mean-field, alpha-omega dynamos, which are
extensively studied both analytically and numerically. Various dynamo
modes may exist, resulting in characteristic configurations of the
global galactic field. Specific dynamo modes may indicate the
influence of an external disturber or of a bar \citep{Ferriere07}.
In galactic disks, the dynamo mode most easily excited is axisymmetric
and has even parity with respect to the midplane (i.e., the horizontal
field component has the same direction on either side of the midplane,
while the vertical component changes sign across the midplane \cite[e.g.,][]{Ruzmaikin88}).  The field is dominated by its toroidal
component, which together with the radial component, forms an
axisymmetric spiral pattern.  Rarely observed bisymmetric fields in
galactic halos could be relics of the initial intergalactic magnetic
field (as suggested by \cite{Fletcheretal11} for the halo of M\,51).
As opposed to flat disks, dynamos in spherical objects -- such as
galactic halos -- are expected to favor the axisymmetric mode with odd
vertical parity, in which the magnetic field crosses the midplane
  continuously at right angles and the horizontal component reverses
direction across the midplane\footnote{Note that the even-parity
  poloidal field is also said to be quadrupolar, and the odd-parity
  poloidal field is also named dipolar, although it generally differs
  from a pure quadrupole and dipole, respectively.}.

In the disk and halo system of the Milky Way, this would result in
mixed-parity modes \citep{SokoloffShukurov90}. Although
\cite{MossSokoloff08} find that in a disk-halo system, usually the
stronger dynamo enslaves the other, resulting in either an even-parity
or an odd-parity magnetic configuration, mixed-parity modes can be
obtained if a moderate Galactic wind is included \citep{Mossetal10}.
The geometry and parity of the halo magnetic field hold the key to its origin.
For instance, an X-shaped field would probably indicate the existence of 
a large-scale Galactic wind. Moreover, even parity of this X-shaped field 
would suggest a wind originating near the Galactic plane and advecting 
the dynamo field of the disk into the halo, while odd parity would suggest 
a wind blowing from the base of the halo and stretching out the (normally 
dipole-like) dynamo field of the halo into an X shape \citep{FerriereTerral14}.

Current observations show a picture generally consistent with dynamo
models, albeit so rough that it is impossible yet to determine how
exactly the Galactic magnetic field is maintained. It is now
well-established that the large-scale magnetic field in the Milky Way
disk is generally aligned with the spiral arms and has a strength of
about $3~\mu$G (see the review by \cite{Haverkorn14}). However, there
is still considerable debate over a few basic properties of the
large-scale field, including its azimuthal structure (i.e., whether it
is axisymmetric, bisymmetric, or neither as concluded by
\cite{Menetal08}), and the number and location of large-scale
reversals in the field direction.  Most recent models based on
multiple observational tracers favor an axisymmetric spiral structure
with a single reversal just inside the Solar circle
\citep{Brownetal07,Sunetal08,VanEcketal11,JanssonFarrar12}. However,
some of the models based on only pulsar data conclude that multiple
reversals must exist \citep{Hanetal06,NotaKatgert10}.  In
principle, large-scale field reversals can be produced by dynamos when
the rotational velocity has a large vertical gradient
\citep{FerriereSchmitt00}, or be a result of early injection of
small-scale fields \citep{Hanaszetal09,Mossetal12} or of
magneto-rotational instability \citep{Machidaetal13}.  Much of the
controversy about the presence of large-scale field reversals in the
Milky Way is fueled by the lack of detection of field reversals in
external galaxies, with the exception of M\,81 which appears to host a
bisymmetric field (see \citet{becketal15} for a detailed discussion). 

Recent Galactic magnetic field models seem to confirm the mixed parity
dynamo modes in disk and halo \citep{Fricketal01,Sunetal08,
  JanssonFarrar12}. This also would explain the apparent
  contradiction between the observed ``butterfly pattern'' of rotation
  measures in the inner Galaxy, indicating a reversal of the azimuthal
  field component across the Galactic plane
  \citep{SimardNormandinKronberg80} and local measurements of field
  direction indicating no reversal of this component across the
  galactic plane \citep{Fricketal01}. The first would be associated
  with the halo field, whereas the second reflects the disk field.
However, characterizing the global halo field structure is made
difficult by the contamination of local structures. As an example, the
Faraday rotation pattern of extragalactic sources recognized as a
large-scale dipolar field in previous studies is strongly influenced
by a local magnetized bubble \citep{Wollebenetal10,
  SunReich10}. Several forms of the halo magnetic field geometry have
been proposed in the literature, including a purely azimuthal
double-toroidal component (e.g., \cite{SunReich10}), a spiral
component (e.g., \cite{Maoetal12}) and an out-of-plane X-shaped
component (e.g., \cite{JanssonFarrar12}) as observed in external
edge-on spiral galaxies and as modeled by
\cite{FerriereTerral14}. Vertical field components are expected
  in the Milky Way, by analogy to nearby spirals
  \citep{becketal15}. However, the vertical component of the field at
  the Galactocentric radius of the Sun is observed to be small
  \citep{Maoetal10}. This suggests that the Milky Way's vertical field
  components are concentrated towards the inner Galaxy, as modeled in
  e.g. \citet{JanssonFarrar12}.

\begin{figure}
\begin{center}
\includegraphics[width=.45\textwidth]{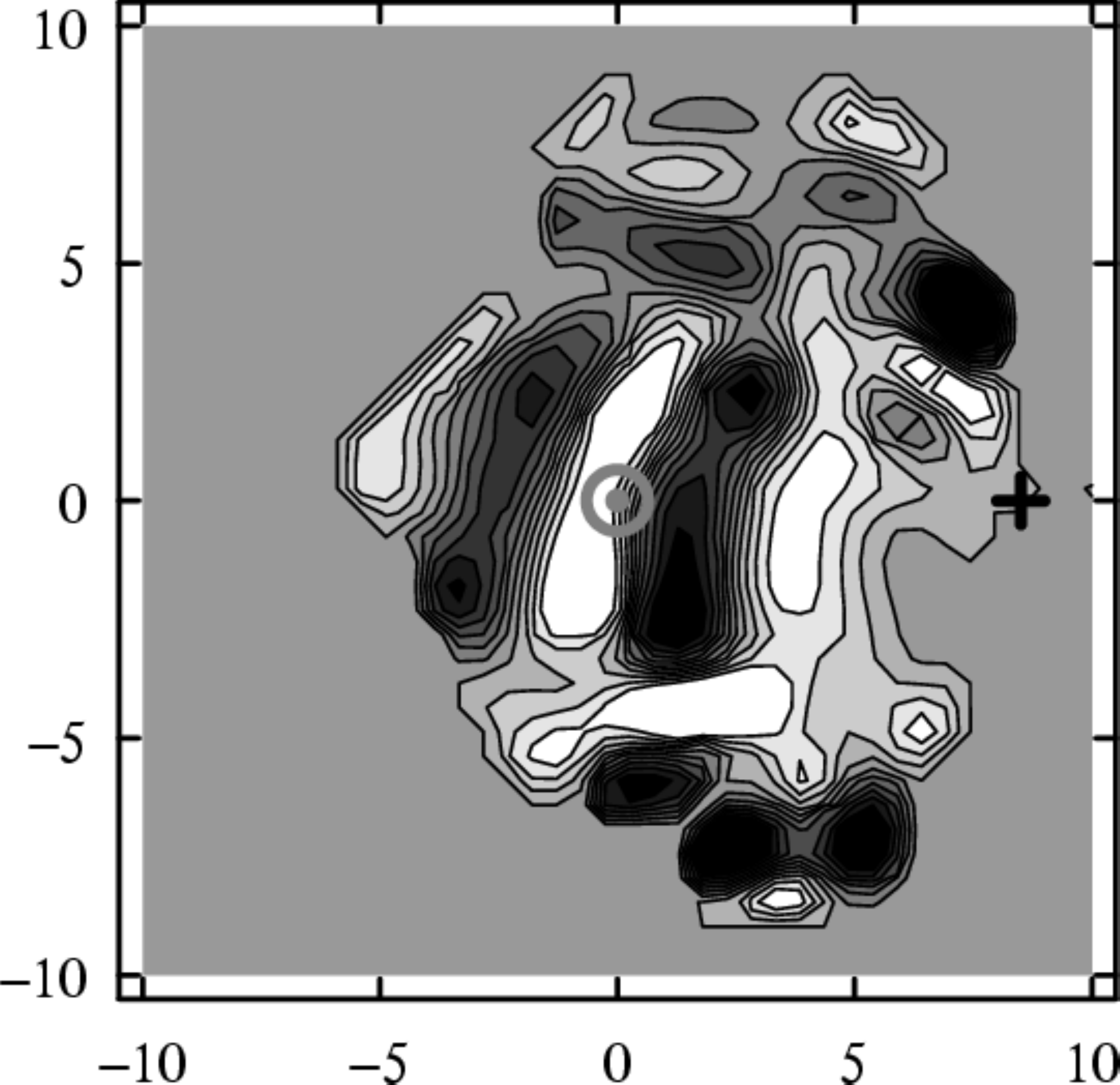}
\includegraphics[width=.45\textwidth]{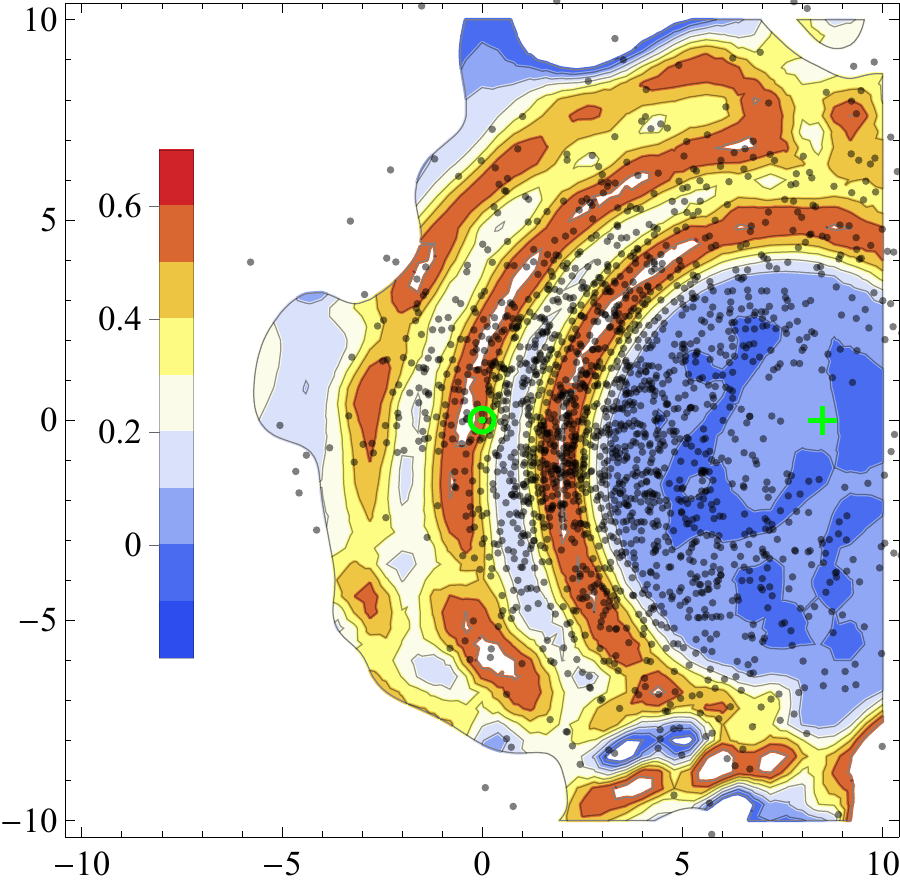}
\caption{Wavelet transforms of a test function containing a Galactic
  electron density model, retrieved from pulsar RMs. The Galactic
  Center is marked with a cross and the Sun is at the center of the
  frames. The axes are in kpc, and in uniformly grey/white regions,
  data is too sparse. Left: current situation, using 323 pulsars
    with known RMs \citep{Stepanovetal02} and a two-arm model. Right:
    the same with a simulated pulsar data set of 2000 pulsars to be
    expected with SKA1, fitting well to a three-arm model. }
\label{f:stepanov02}
\end{center}
\end{figure}

The observations on which these magnetic field models are based
consist of RM Grids of extragalactic sources and pulsars, and/or
diffuse synchrotron emission. The models are hampered by low density
of polarized sources, uncertain distance estimates of pulsars, and
contamination by foreground structures such as supernova remnants
(e.g.\ the radio loops).  A good example of this is the wavelet
analysis of pulsar data to determine the large-scale structure of the
Galactic magnetic field \citep{Fricketal01}, which shows good results
in the Solar neighborhood, but cannot determine the global field
structure due to insufficient data points \citep{Stepanovetal02}. This
is illustrated in the left plot of Figure~\ref{f:stepanov02},
which shows a wavelet transform of a test function representing
  Galactic electron density obtained from an irregular grid of
pulsar data \citep{Stepanovetal02}. The plot demonstrates the method's
unique capability to trace out magnetic spiral arms, but also that
this is currently only possible in the Solar vicinity.

\subsection{Galactic turbulence}

Both ionized and neutral phases of the Galactic ISM are turbulent,
usually displaying a Kolmogorov-like power spectrum in density across
a wide range of length scales
\citep{ElmegreenScalo04,1995ApJ...443..209A}. Determining the power
spectrum of the magnetic field is much more difficult: reliable
  estimates only exist for related quantities like RM. Observed
spectral indices\footnote{Usually, power in RM fluctuations is
  measured by structure functions (as explained below rather than
  power spectra, the slope of which is directly related to power
  spectral index.} of RM are generally much flatter than Kolmogorov
\citep{Haverkornetal08}.

  Besides the spectral index, magnetohydrodynamic (MHD) turbulence can
  be characterized by several fundamental parameters including scales
  of energy injection and dissipation, the rms Mach number of flow
  speed, intermittency, and the plasma beta (the pressure ratio of the
  gas to the magnetic field). The search for such characteristics is
  important to understand the dynamical and thermal states of the
  ISM. In addition, numerical modeling of turbulence is essential to
  understand the physics behind the observations of turbulence
  parameters. E.g., modeling revealed that known turbulent structures
  are not enough to explain observed structure
    functions\footnote{The structure function of order $n$, of a
    quantity $f$ as a function of scale $\mathbf{\delta x}$ is defined as
    SF$_f(\mathbf{\delta x}) = \langle |f(\mathbf{x}) - f(\mathbf{x}+\mathbf{\delta x})|^n \rangle_{\mathbf{x}}$,
    where $\langle \rangle_{\mathbf{x}}$ denotes averaging over all positions
    $x$. The second-order structure function ($n=2$) is most used and
    usually meant when just ``structure function'' is used.} (SFs) of RM toward high Galactic latitudes
    \citep{2013ApJ...767..150A}. A possible cause of the discrepancy
    would be magnetic fields emerging from the disk to the halo due to
    the magneto-rotational and Parker instabilities
    \citep{Machidaetal13}.

\begin{figure}
\includegraphics[width=\textwidth]{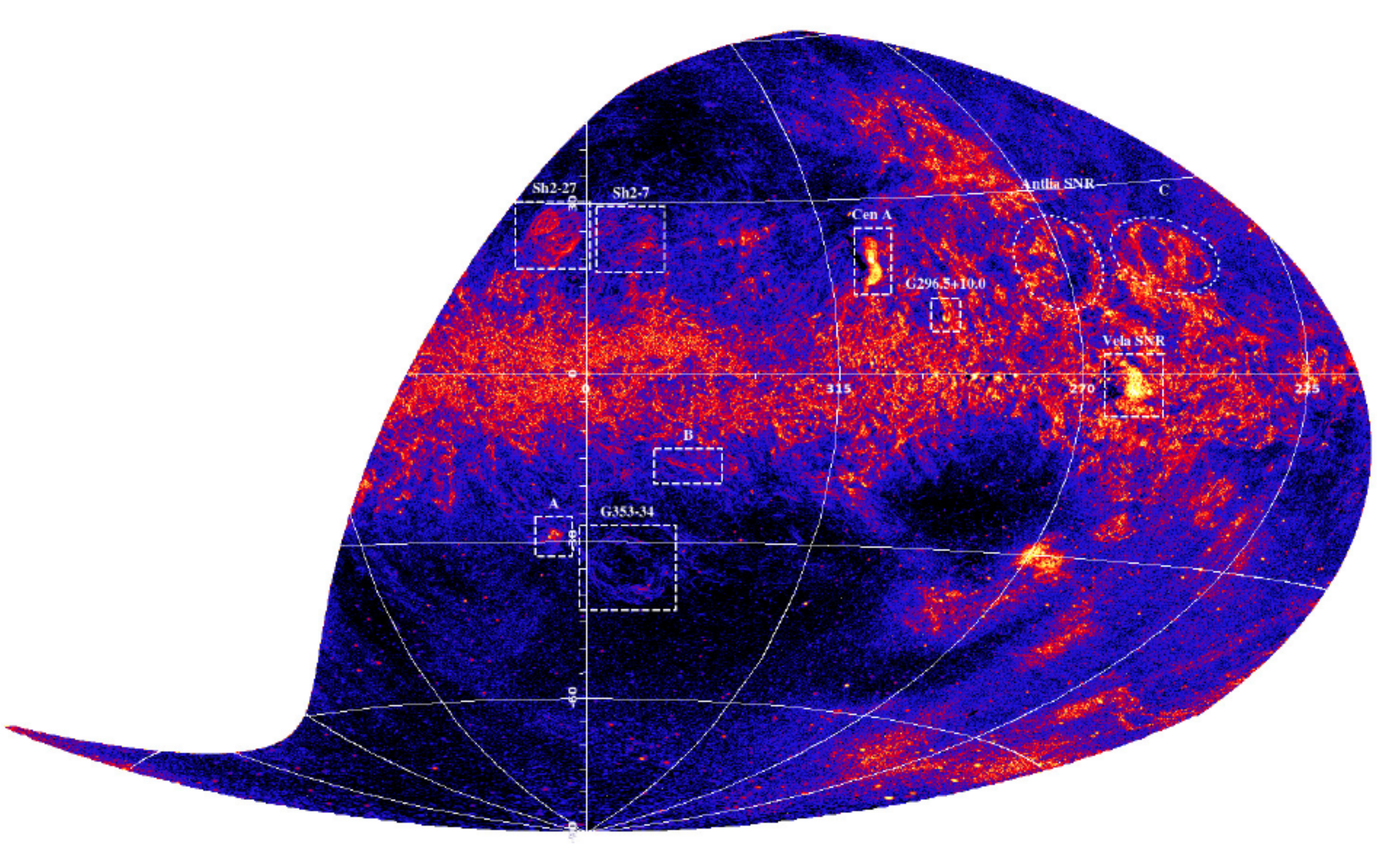}
\caption{Gradient of the polarization vector over the entire southern
  sky \citep{Iacobellietal14}. Analysis of the probability distribution
function of the polarization gradient shows that interstellar
turbulence is transonic in the regions studied.}
\label{f:iacobelli14}
\end{figure}

The Mach number of interstellar turbulence can be determined through
polarization data, by way of the gradient of the polarization vector
\citep{2011Natur.478..214G}. Application of this method on a small test
region, combined with simulations \citep{Burkhartetal12}, reveals that
interstellar turbulence is transonic with $M\lesssim
2$. \cite{Iacobellietal14} showed that this Mach number value is
roughly consistent over the entire (Southern) sky, as can be derived
from the polarization gradient shown in Figure~\ref{f:iacobelli14}.

Measured maximum scales of fluctuations in the magneto-ionized medium
are likely to be connected to the scale of energy injection in the
interstellar gas. Energetics imply that the main energy sources likely
to drive interstellar turbulence are supernova remnants.  Early
observational studies using correlation functions corroborated this,
concluding that the outer scale of turbulence must be $O(100)$~pc
\citep{LazaryanShutenkov90}. Usually, these maximum scales are probed
by second-order SFs of RM, to be obtained from measurements of RM
Grids. However, recent observations indicate maximum scales of
  fluctuations smaller than $\sim 10$~pc \citep[in spiral arms
  only,][]{Haverkornetal08}, of a few parsecs from anisotropies in TeV
  cosmic ray nuclei \citep{Malkovetal10}, and similar from analysis of
  synchrotron fluctuations \citep{Iacobellietal13}. These small scales
do not necessarily indicate the maximum scale of turbulence, but only
show that larger-scale fluctuations in the magneto-ionic medium are
not observed. Turbulence may still be present on scales up to
$\sim$100~pc, but this structure may be masked by small discrete
sources \citep{Haverkornetal06b}, strong small-scale fluctuations in
electron density, dominating shock waves in spiral arms, etc. The SF at higher
Galactic latitudes has a flatter slope in $>1^\circ$ scales
\citep{2011ApJ...726....4S}. These authors claim that the steep slope
of SFs at low latitude could be affected by the local turbulent ISM.

Some evidence exists that interstellar turbulence in dense gas is
intermittent, i.e.\ contains regions with intense turbulence and
regions with no turbulence, giving long tails in the distribution of
the observable such as density or velocity
\citep{ElmegreenScalo04}. This can be detected using higher order
SFs. However, in the diffuse ionized gas, current SFs contain far too
few sources to be able to construct meaningful higher order SFs.

\subsection{Magnetic fields in discrete objects}

\paragraph{Star forming regions} The launching mechanism for jets in
general has been proposed to be universal (e.g. \cite{Livio11}),
crossing orders of magnitude in terms of source energetics and
environmental conditions. However, our observational understanding of
jets in Young Stellar Objects (YSOs) is almost diametrically opposed
to that of their higher energy analogues such as AGN, GRBs and X-ray
binaries. While characteristic quantities such as density, temperature
and ionization fraction can be well constrained for YSOs from
(primarily) line emission data, the strength and orientation of the
magnetic field in these systems is still very poorly known
\citep{Ray09}. Observational evidence for these magnetic fields is
generally indirect and model dependent: jet rotation, cushioning
effects on jet shocks, etc. Although non-thermal radio emission can
provide a direct measure of magnetic field strength and configuration,
measurements of these properties are very rare \citep{Rayetal97,
  CarrascoGonzalezetal10, Ainsworthetal14}. This is because the
associated non-thermal emission is faint ($\sim\mu$Jy), and because
radio studies of star forming regions are predominantly total
intensity, cm-wave experiments focusing on thermal radio emission from
ionized gas.

In spite of this, the role of the magnetic field in jet launching is
expected to be inextricably linked to the active accretion of the
central object from its gaseous disk and surrounding molecular cloud,
through which Galactic magnetic field lines are known to be threaded
and where the influence of the magnetized jet may provide turbulent
support against collapse. The centrifugal acceleration of material
along these field lines is widely accepted to be closely connected to
the launching of YSO jets through MHD processes, as well as the
collimation of those jets through a "winding-up" of the magnetic field
lines anchored in the rotating disk leading to the creation of hoop
stresses in the resulting helical field configuration
\citep{BlandfordPayne82,Ferreiraetal06}. Furthermore, the magnetic
fields in these systems are crucial not only for understanding the
star formation process but, through the acceleration of relativistic
particles in jet shocks, also act as cosmic-ray factories,
contributing to the GeV excess in the Galaxy
\citep{OrlandoStrong13}.\footnote{For SKA measurements of magnetic
  fields in molecular and cold gas, linked to the early phases of star
  formation, we refer the reader to \citet{Robishawetal15}.}

\paragraph{Supernova remnants} Magnetic fields in supernova remnants
(SNRs) can be enhanced to mG strengths. Magnetic field strengths in
young SNRs are generally radially directed, whereas old SNRs have
mostly tangential fields, although turbulent fields play a significant
role (for a review see \cite{Reynoldsetal12}). Apart from the
importance of magnetic fields for the evolution of SNRs themselves,
these fields are crucial for acceleration of Galactic cosmic rays
\citep{Blasi13}. SNRs are also believed to be the main drivers of
interstellar turbulence.

There is some evidence from both observations
  \citep{Gaensler98,Uyanikeretal02,KothesBrown09} and numerical
  simulations \citep{Stiletal09} that magnetic fields in SNRs carry
  the signature of larger-scale Galactic magnetic fields which they
  are expanding in. Similarly, in some SNRs, there is evidence for
  toroidal fields, interpreted as the product of the stellar wind of
  the progenitor, swept up by the blast wave
  \citep{Uyanikeretal02,HarveySmithetal10}. Some massive stars exhibit
  non-thermal radio emission, so magnetic fields must occasionally be
  present in the winds. Nevertheless, \cite{Delachevrotiereetal14}
  failed to find optical evidence of magnetic fields in 11 WR stars
  (by searching for circular polarization). Only a large sample of SNR
  magnetic fields will shed light on the relationship between magnetic
  fields in SNRs and stellar and/or Galactic magnetic fields.

Measurements of magnetic fields in Pulsar Wind Nebulae (PWNe) are
scarce and biased towards young, bright PWNe. However, an intriguing
example of the importance of observing PWN magnetic fields is given by
\citet{Kothesetal08}, who showed that the older PWN DA~495 has a
dipole-like field, giving an estimate of the spin axis of the putative
pulsar. The PWN magnetic field configuration is best studied through
depolarization and Faraday tomography of the PWN emission itself, but
probing through RMs of background sources can contribute. For more
discussion on PWN science with the SKA, see \cite{Gelfand15}.

For a discussion on other SNR science with the SKA, see \citet{wang15}.

\paragraph{HII Regions}
Typical magnetic field strengths in HII regions are a few to about
$12~\mu$G \citep{Sunetal07,Gaoetal10,HarveySmithetal11},
obtained from observations of the Faraday rotation of background
Galactic synchrotron emission. Magnetic field measurements in HII
regions over a large range of densities should be obtained to
understand the dynamic role that magnetic fields play. This can be
done using a dense RM Grid. Str\"omgren spheres tend to be so dense
and so turbulent that they are Faraday thick at 1~GHz, so that
emission from the objects themselves can only be probed at higher
frequencies, where the regions become Faraday thin.

Magnetic fields in HII regions also seem to retain knowledge
  about larger-scale Galactic magnetic fields
  \citep{HanZhang07,HarveySmithetal11}. In particular, this should be
  true of Planetary nebulae (PNe), which do not have an intrinsic
  magnetic field. However, some have recently been shown to have a
  polarization signature \citep{Ransometal08,Ransometal10}.
  Therefore, mapping out magnetic field directions in PNe gives an
  independent probe of the large-scale field. Extensive catalogues
  exist of PNe, down to arcsecond size objects. Measurements of the
  Faraday rotation in the ionized shell can be coupled with optical
  observations which can supply data on the distance, the size of the
  ionized region, and the electron density. This can lead to the
  determination of unambiguous magnetic field values all over the
  Galaxy.

\paragraph{Faraday screens}

Faraday screens are structures observed through their polarization
properties \citep{Grayetal98,Haverkornetal03,WollebenReich04}, which
are difficult to identify as physical Galactic objects from turbulent
structures. These objects are invisible or at best very faint in the
continuum, but visible in polarized emission and RM. Some of them can
be described as magnetic bubbles with strong regular fields and low
thermal electron density, when their magnetic pressure exceeds thermal
pressure. At 4.8~GHz, Faraday screens with high RMs of the order of
200 rad~m$^{-2}$ were identified in the Galactic plane in polarized
emission \citep{Sunetal07,Gaoetal10}, whereas lower-frequency
  data reveal screens of a few to a few tens of rad~m$^{-2}$. Many
  more Faraday screens are expected to have lower RMs, but these are
  easily be masked by small-scale polarization from the magnetized
  turbulent ISM.  Some of them may be linked to faded PWNe or old HII
regions leaving expanding magnetic bubbles \citep{Iacobellietal13},
but in general their nature and origin are still enigmatic.

\subsection{The Galactic Center}
\label{s:1gc}

The Galactic Center (GC) is important -- aside from its intrinsic
interest -- because it is, by definition, the closest example of a
galactic nucleus that we possess. It has a complex magnetic field
structure in interaction with the gas and relativistic particles close
to the GC (see \cite{Ferriere09} for a review). The GC magnetic
  field is most likely a result of dynamo action, either a local
  dynamo or the dynamo maintaining the Galactic
  magnetic field, which may require a central galactic wind to
  stabilize \citep{Mossetal10}. \citet{Crockeretal11} showed evidence
  that a GC wind is indeed present. Therefore, measuring and
  understanding magnetic fields in the GC is also a key component in
  understanding Galactic magnetic fields.

Large uncertainties still exist about the field strength. Estimates
are $\sim(6-22)~\mu$G (minimum energy), $\gtrsim (50-120)~\mu$G
(synchrotron spectral break, \citet{Crocker10,Crocker13}) or
$\sim1$~mG (pressure balance) in the general intercloud
medium. Measurements of the RM and dispersion measure (DM) of the
magnetar J1745-2900 indicate that the magnetic field at $\sim~0.1$~pc
from Sgr A* could be as high several mG \citep{Eatoughetal13}. Zeeman
measurements and submm polarimetry in dense clouds show field
strengths of 0.1~mG to a few~mG
\citep[e.g.,][]{Killeenetal92,Crutcheretal96,Chussetal03}.

The configuration and structure of the GC magnetic field are also
significantly unclear. \citet{Novaketal03} and \citet{Law2011} used a RM Grid to
obtain a magnetic field geometry organized on $\sim300$~pc
scales, centered some 50~pc west of the dynamical center of the
Galaxy. They interpret polarization in radio synchrotron filaments as
a poloidal (vertical) field in the diffuse ISM. This is in conflict with the findings of
\cite{Roy2005,Roy2008}, who obtained results -- based on an RM Grid of
64 background sources -- consistent with a bi-symmetric spiral
magnetic field in the Galaxy or with a field oriented along the
central bar of the Galaxy, with a field strength of $\sim20$~$\mu$G.

Unanswered questions also remain about how the recently-discovered
Fermi Bubbles are related to the central regions of our Galaxy: are
they the result of electron-dominated processes \citep{Su2010} from
close to the $\sim4.1\times10^6~M_\odot$ black hole at the center, or
are they formed by proton dominated objects \citep{Crocker2011}
arising from the stellar activity in the central $\sim200$~pc?
Evidence for the latter seems to have emerged from polarized radio
counterparts to the Fermi bubbles detected in the S-PASS survey
\citep{Carrettietal13}.  The magnetic field structure is a critical
part to the resolution of this question, since the field
  structure is likely intimately tied to both the maintenance of the
  dynamo and the formation of Bubble-like structures in galaxies
  through the establishment of winds.

\section{Galactic Magnetism Science with the Square Kilometre Array}
\label{s:ska}

\subsection{Methods}
\label{s:methods}

Although the Galactic Science described in this chapter is very
diverse, the science goals can be reached using two main observational
strategies, viz.\ the RM Grid and Faraday tomography with broadband
polarimetric observations of diffuse Galactic radio emission. We
describe these methods in this Section, and then explain how we can
reach our science goals with these observations in the SKA1 phase,
SKA1 Early Science and in SKA2.

\subsubsection{The RM Grid}
\label{s:methods:grid}

The {\em RM Grid} as described in the 2004 SKA Science Case
\citep{BeckGaensler04} continues to be the highest priority for the
Cosmic Magnetism Working Group (see \citet{johnstonhollitt15}). The RM
Grid possible with SKA1 is based on the summary of the Cosmic
Magnetism Science Working Group Assessment Workshop
\citep{SKAmemo}. These observations focus on a relatively shallow
spectro-polarimetric survey at $\sim 1$~GHz over a broad frequency
coverage, enabling a vastly improved RM Grid. Currently, about 40,000
extragalactic background sources with measured RMs are available in
the RM Grid, mostly based on \cite{Tayloretal09}.
A SKA1-SUR Band 2 survey with $2\mu$Jy sensitivity (4 hours per
pointing with 500 MHz bandwidth) may provide RM Grids of 300-1000
sources/deg$^2$, i.e.\ an improvement of $2-3$ orders of magnitude
to the current status.

Pulsar RM surveys also offer a vast improvement in data
quality. \cite{Smitsetal11SKAmemo} estimate that SKA1-LOW will provide
$9,000-10,000$ new pulsar detections, including their RMs, whereas
SKA1-MID will discover about $12,000-13,000$. Recent estimates
  predict that $30,000$ new pulsars will be discovered with the SKA,
  50\% of which will already be found with SKA1
  \citep{KramerStappers15}. These pulsar measurements become
exceptionally valuable for modeling the Galactic magnetic field when
reliable distance estimates are available. For this goal, the step to
SKA2 will be crucial: in SKA2, timing parallaxes should allow distance
measurements with an accuracy of about 5\% up to a maximum distance of
30~kpc, which means throughout the Milky Way
\citep{Smitsetal11}. See \citet{hanetal15} for more detailed
discussion of the use of pulsars
to investigate the magneto-ionic medium of the Milky Way.

\subsubsection{Faraday tomography}

In the past few years, Faraday tomography\footnote{For an extended
  explanation, see \citet{healdetal15}.} has started to provide
new views into the structure of the Galactic magnetic field, albeit
still incomplete and puzzling. The few studies available for various
sightlines all find that the path length is filled with a small number
of discrete synchrotron emission regions, interspersed with
Faraday-rotating screens \citep{Schnitzeleretal07,
  Brentjens11,DeBruynPizzo14, Jelicetal14}. This hints at the
complexity of the magnetic field structure, but is at least partially
due to the insensitivity of current observations to Faraday-thick
structures.

The greatest leap in knowledge about Galactic magnetism may well lie
in the opportunities that Faraday tomography will give. Long
wavelengths are required for a high maximum detectable Faraday
thickness of $\Delta\phi_{\mbox{max}} = \pi/\lambda_{min}^2$, while
broad frequency coverage provides a high Faraday depth resolution
$\delta\phi = 2\sqrt{3}/\Delta\lambda^2$ \citep{BrentjensDeBruyn05}.
Typical Faraday depths in the tomography studies above are a few to a
few tens of rad~m$^{-2}$, while objects like SNRs or PWNe show Faraday
depths up to a few hundred rad~m$^{-2}$. Therefore, a Faraday
tomography survey needs high Faraday depth resolution and capability
to observe Faraday thick structures (up to $\sim 100$~rad~m$^2$),
necessitating broad frequency coverage at relatively low frequencies
($< 1$~GHz).

The broad frequency coverage needed for Faraday tomography is a key
aspect for the broadband polarimetry project described by
\cite{Gaensleretal15}. Even though their science goals focus on
Faraday tomography of extragalactic radio sources themselves, the data
can be taken commensally for the two projects. The parameters needed
for their survey match well with the Galactic science goals described
here.

\subsection{Prospects for SKA1-Early Science}

For the SKA1 Early Science phase, we intend to use the results from a
combined all-sky polarimetric survey with SKA1-SUR Band 2 with the
ASKAP-POSSUM survey \citep{Gaensleretal10} as proposed by
\citet{Gaensleretal15}. These combined data would have continuous
frequency coverage from 700~MHz to 1500~MHz, which translates into a
maximum detectable Faraday thickness $\Delta\phi_{\mbox{max}} \approx
80$~rad~m$^{-2}$ and a Faraday depth resolution $\delta\phi =
24$~rad~m$^{-2}$. 

For studies of the ordered and random components of the Galactic
  magnetic field, the Faraday depth resolution of this Early Science
survey is too low, although analysis techniques like QU-fitting
\citep{OSullivanetal12} can increase the Faraday depth resolution
significantly.

However, this survey is well suited to study discrete objects like
SNRs, which have higher observed RMs. In research on SNRs and
other discrete objects, the SKA Pathfinders will already make big
advances, given their ability to image SNRs at the other side of the
Galaxy ($10''$ resolution at a distance of 15 kpc gives a physical
resolution of 1~pc). This will allow some sense of magnetic field
strength in these objects.

However, the proposed combined SKA1 Early Science polarimetric survey
will be able to {\em map} magnetic field strengths in SNRs. This will
increase the number of SNRs with known magnetic field structure by
orders of magnitude, finally making statistical studies of SNR
magnetic fields possible.  Faraday depths within a SNR have been found
to vary by $100-200$~rad~m$^{-2}$
\citep{Uyanikeretal02,HarveySmithetal10}. If these Faraday thick
  components are uniform, this survey will not be able to detect
  them. However, the Faraday depth maps show ample small-scale
  structure, making it plausible that a large part of the Faraday
  depth can be detected. Any missing large-scale Faraday depth may be
  estimated from the RM Grid. Assuming that we want to probe such a
SNR in 10 Faraday depth slices, the proposed survey's sensitivity of
$\sim 7~\mu$Jy~bm$^{-1}$ at a resolution of $10^{\prime\prime}$ will
increase the number of SNRs for which we have magnetic field
measurements from a few to a few hundred\footnote{Compare to current
  capabilities of e.g.\ the Canadian Galactic Plane Survey SNR catalog
  \citep{Kothesetal06}, where 24 SNRs were detected in polarization
  (single channel) at 1420~MHz, over 110$^{\circ}$ of the outer
  Galactic plane, with a sensitivity of $\sim 5$mJy~bm$^{-1}$.}.

\subsection{Prospects for SKA1}

With the increased sensitivity of SKA1 with respect to the Early
Science phase, weaker magnetic fields will become visible, such as the
interstellar magnetic fields in the Galactic plane and halo, using
both broadband polarimetry of diffuse emission and Faraday tomography.
Again we suggest to commensally use the proposed Broadband Polarimetry
Survey for SKA1 \citep{Gaensleretal15}, which will have full frequency
coverage over a range of $650-1750$~MHz at 1$^{\prime\prime}$
resolution (ten times the resolution at Early Science) over the whole
sky down to a sensitivity of $\sim 5~\mu$Jy~bm$^{-1}$. The high
resolution and sensitivity can be used to unveil the non-thermal
emission in star forming regions, and probe SNRs and HII regions to
larger distances.

However, detailed mapping of the interstellar magnetic field will need
coverage at lower frequencies to obtain sufficient Faraday depth
resolution. Assuming that Band~1 will be available for SKA1-SUR, an
all-sky survey in the $350-650$~MHz range to complement the SKA1-Early
Science survey is needed to provide a Faraday depth resolution of
$\delta\phi = 5$~rad~m$^{-2}$. As detailed in Figure~7 in
\cite{Braun13}, a $10^{\prime\prime}$-resolution all-sky survey down
to $\sim 7.6~\mu$Jy~bm$^{-1}$ across $350-650$~MHz would imply 2 full
years of observing time. Therefore, obtaining these low frequencies
down to the sensitivity of the proposed SKA1-SUR Band 2 survey would
be better as a SKA2 project (see below).

\paragraph{Galactic magnetic fields}
In order to assess the possibilities of observing Galactic turbulent
magnetic fields with SKA, simulations of high-resolution SKA
observations were made of total intensity and polarized Galactic
emission in various directions of the sky within the SKADS program
\citep{SunReich09}.  Based on realistic Galactic 3D-models of the
distribution of thermal electrons, cosmic-ray electrons, and magnetic
fields \citep{Sunetal08}, 1.6$^{\prime\prime}$ resolution maps at
1.4~GHz for total and polarized intensities and RMs were calculated,
similar to the proposed survey above.  The extent of diffuse
structures depends on the outer scale of the assumed Kolmogorov
spectrum for the turbulent magnetic field, the length of the
line-of-sight through the Galaxy and also the regular magnetic
field. These studies show that even at high latitudes, Galactic total
and polarized emission is highly structured at arcsec resolution and
must be separated from sensitive SKA maps of extended extragalactic
objects in an appropriate way.

Studies of magneto-ionized turbulence through SFs will benefit greatly
from a tight RM Grid as provided by SKA1. A grid with 1000 sources per
square degree could unveil the second-order SF down to arcmin
scales. The survey capability allows us to make the first detailed
all-sky map of the energy injection scale, in order to reveal the
origins of driving forces. Moreover, a large number of sources with
small RM errors (e.g. $<1$~${\rm rad/m^2}$) allows us to derive
higher-order SFs. Higher-order SFs provide further valuable
information of turbulence, e.g.\ the trend of SF slopes to higher
order depends on the characteristics of turbulence. 

SKA1 pulsar studies will detect many more pulsars, including more
  DM measurements and reliable distances. The right hand plot in
  figure~\ref{f:stepanov02} shows the wavelet transform result of a
  test function representing electron density is obtained using a
  simulated pulsar data set of 2000 pulsars expected to be detected by
  SKA1 (R. Smits, priv. comm.). It is obvious from this plot that this
method promises major progress in the unraveling of the large-scale
structure of the Galactic magnetic field.

\paragraph{Supernova remnants}

With SKA1, the RM Grid will provide one source every few
arcminutes. This means that we will be able to roughly probe the
magnetic field of virtually every SNR in the Galaxy using the grid.
Faraday tomography with SKA1-LOW/SKA1-SUR Band 1 will delineate
magnetic fields in the SNR shell in 3D in pointed observations of
individual target sources. Statistical samples of magnetic field
structure in HII regions, PWNe, Planetary Nebulae can be built up with
these data. Also, SKA1-LOW will enable a full magnetic field survey of
nearby SNRs, but not until the enhanced sensitivity of SKA2 will we be
able to probe magnetic fields in SNRs throughout the Milky Way in this
way.

\paragraph{Star forming regions}

The high resolution of the SKA1 survey mentioned above can be used to
unveil the non-thermal emission in star forming regions, reconciling
the wider Galactic magnetic field with MHD models of star formation,
protostellar accretion, disk coupling and jet launching. Emission is
likely to be present at $\sim50~\mu$Jy at 1000~MHz, with a spectral
index of $-1$ \citep{Ainsworthetal14}, i.e.\ a $10\sigma$
detection. SKA1-MID can be used for follow-up to obtain the
spectrum. Using the SKA1-MID sensitivity of $63~\mu$Jy~hr$^{-1/2}$
over a 100~kHz band, a spectrum with a $3\sigma$ detection at 1~GHz
over 10~MHz channels can be obtained in 30~mins per source.
To date, two YSO jets have been detected in pointed observations
\citep{Rayetal97,CarrascoGonzalezetal10}, making it impossible to
estimate reliable detection rates at the moment. We assume that we
will be able to inspect the nearest star formation regions in total
intensity with SKA1. Even a detection of a handful of YSO jets will
make a large impact on the source sample.

\paragraph{Galactic Center}
The dense SKA1 RM Grid will allow unprecedented modeling of the
structure and strength of the magnetic field in the GC,
resolving the tension in the findings described in
Section~\ref{s:1gc}.
In particular, detection of 15,000 new pulsars with measured RMs will
allow us to understand the overall structure of the GC magnetic field
(i.e., is it toroidal?  Poloidal?) in a comprehensive manner, and how
this connects to the overall Galactic magnetic field
structure. Higher-order SFs will become possible due
to the high source density in the SKA RM Grid, which will enable
determination of hereto elusive properties of interstellar turbulence
such as intermittency.  The RM Grid will identify Faraday
screens and clarify their role in the ISM and hopefully also their
origin.

\subsection{Prospects for SKA2}

A tenfold increase in sensitivity of SKA2 at frequencies from 350~MHz
to 24~GHz will allow extension of the all-sky broadband polarimetric
survey at high sensitivity down to lower frequencies. This will make a
Faraday tomography all-sky map of magnetized turbulence possible,
allowing not only to fully characterize these Galactic magnetic
fields, but also to describe in detail polarized foregrounds for
various cosmological research areas such as the Epoch of
Reionization. Also, this increase in sensitivity for SKA2 will enable
larger samples of YSO's, making statistical studies possible, and
detection of linear polarization, adding information on the
orientation and ordering of the field. Finally, Faraday tomography for
SNRs throughout the Milky Way can be done, giving a complete sample of
3D magnetic field structure in SNRs throughout the Milky Way.

For the RM Grid, SKA2 will provide us with accurate distances for just
about any detectable pulsar in the Milky Way. The large increase of
reliable pulsar RMs and distances will finally allow resolution of the
issue of number and location of large-scale reversals in the Milky
Way, thanks to the accurate parallax distances measured throughout a
major part of the Galaxy. Wavelet analysis as in \cite{Stepanovetal02}
will be possible (a) for the entire Galaxy, and (b) down to scales of
about 100~pc, which is comparable to some current estimates of
the turbulence injection scale.

Additionally, the SKA2 phase will also allow a high angular and
$\lambda^2$-space resolution observation of non-thermal radio
filaments in the GC such as the Radio Arc, which
possesses RMs of $\sim-3\times10^3$~rad~m$^2$
\citep{YZ1987} -- some of the highest known in our Galaxy -- which will
allow a detailed modeling of their origin, as well as discovery of
numerous others at lower surface brightness.

\bibliographystyle{apj}

\end{document}